\documentclass[prl,twocolumn]{revtex4}
\usepackage{graphicx}
\usepackage{color}
\usepackage{amsmath}

\usepackage{epstopdf} \DeclareGraphicsExtensions{.eps, .pdf}

\begin{document}

\def\K{{\bf{K}}}
\def\Q{{\bf{Q}}}
\def\X{{\bf{X}}}
\def\Gbar{\bar{G}}
\def\tk{\tilde{\bf{k}}}
\def\k{{\bf{k}}}
\def\q{{\bf{q}}}
\def\x{{\bf{x}}}
\def\y{{\bf{y}}}

\title{Quantum Critical Point at Finite Doping in the 2D Hubbard Model: A
Dynamical Cluster Quantum Monte Carlo Study}
\author{N.\ S.\ Vidhyadhiraja}
\email{raja@jncasr.ac.in}
\affiliation{Theoretical Sciences Unit,\\Jawaharlal Nehru Centre forAdvanced Scientific Research,\\Jakkur, Bangalore 560064, India.}
\author{A.\ Macridin}
\affiliation{University of Cincinnati, Cincinnati, Ohio, 45221, USA}
\author{C.\ Sen}
\affiliation{University of Cincinnati, Cincinnati, Ohio, 45221, USA}
\author{M.\ Jarrell}
\email{jarrell@physics.uc.edu}
\affiliation{University of Cincinnati, Cincinnati, Ohio, 45221, USA}
\author{Michael Ma}
\affiliation{University of Cincinnati, Cincinnati, Ohio, 45221, USA}

\date{\today}

\begin{abstract}
We explore the Matsubara quasiparticle fraction and the pseudogap of the 
two-dimensional Hubbard model with the dynamical cluster quantum Monte Carlo 
method.  The character of the quasiparticle fraction changes from non-Fermi 
liquid, to marginal Fermi liquid to Fermi liquid as a function of doping,
indicating the presence of a quantum critical point separating non-Fermi
liquid from Fermi liquid character.  Marginal Fermi liquid character is
found at low temperatures at a very narrow range of doping where the 
single-particle density of states is also symmetric.  At higher doping 
the character of the quasiparticle fraction is seen to cross over from 
Fermi Liquid to Marginal Fermi liquid as the temperature increases.

\end{abstract}

\maketitle


\paragraph*{Introduction-} The unusual properties of the hole doped cuprate 
phase diagram, including a pseudogap (PG) at low doping, and unusual metallic 
behavior at higher doping have lead many investigators to propose that there 
is a quantum critical point in the cuprate phase diagram at optimal doping.  
Different investigators argue that the PG is related with  the establishment 
of order~\cite{s_chakravarty_01, Varma, Kivelson, Voijta, Sachdev}, where  
the optimal doping is in the proximity of the quantum critical point (QCP) 
associated with this order~\cite{Voijta}.  Other investigators argue
that the QCP is located at the transition from a non-Fermi liquid (NFL) to 
Fermi liquid (FL) ground state without the establishment of order in PG 
region\cite{s_chakraborty_08}.  

In previous work employing cluster extensions of the Dynamical Mean Field, 
a PG region was found at low doping in the two-dimensional Hubbard model.  
It is characterized by a narrow gap-like feature in the single-particle 
density of states, a suppression of the spin susceptibility, and a self 
energy of NFL
character\cite{mark_phase,jarrell:dca,alex_PG,david_PG,choy,tremblay}.  

In this paper, we employ significantly larger clusters than most previous
studies, which affords us greater momentum resolution.  We investigate the 
single particle properties of the two-dimensional Hubbard model with the 
dynamical cluster approximation (DCA)~\cite{hettler:dca,maier:rev}. We find 
further evidence for a QCP and are able to determine its character as the 
terminus of a Marginal Fermi Liquid (MFL) region,  which separates a NFL 
PG region at low doping from a FL region at high doping. We present a 
comparative discussion of a few existing scenarios for quantum criticality 
in the context of our results.


\paragraph*{Formalism-}
\label{sec:formalism}
We consider a 2D Hubbard Hamiltonian
\begin{equation}
\label{eq:hubbard}
H=-\sum_{\langle ij\rangle\sigma}t(c^{\dagger}_{{i}\sigma}c^{\phantom{\dagger}}_{{j}\sigma}+h.c.)+
\epsilon\sum_{{i}\sigma}n_{{i}\sigma}+U\sum_{i}n_{{i}
\uparrow}n_{{i}\downarrow}
\end{equation}
where $t$ is the hopping matrix, $c^{\dagger}_{{i}\sigma}(c_{{i}\sigma})$ 
is the creation (annihilation) operator for electrons on site ${i}$ with 
spin $\sigma$, and $U$ is the on-site Coulomb repulsion which is taken to
be three quarters of the bandwidth $W(=8t)$. The hopping $t$ is restricted
to nearest neighbors $\langle ij\rangle$.

We employ the DCA with a quantum Monte Carlo algorithm as the cluster solver. 
The DCA is a cluster mean-field theory which maps the original lattice model 
onto a periodic cluster of size $N_c=L_c^2$ embedded in a self-consistent 
host.  Spatial correlations up to a range $L_c$ are treated explicitly, 
while those at longer length scales are described at the mean-field level.   
However the correlations in time, essential for quantum criticality,  are 
treated explicitly for all cluster sizes.  To solve the cluster problem we 
use the Hirsch-Fye quantum Monte Carlo method\cite{j_hirsch_86a,jarrell:dca} 
and employ the maximum entropy method\cite{jarrell:mem} to calculate the 
real frequency spectra.

A number of the normal state anomalies in the cuprates are describable by 
a MFL in proximity to a quantum critical point at finite doping\cite{Varma}.  
The imaginary part of the MFL self energy has the form
\begin{equation}
\Sigma_{MFL}''(\omega) = - \alpha \max\left( \left|\omega\right|,T\right)\,.
\label{eq:SigmappMFL}
\end{equation}
In contrast the  imaginary part of the FL self energy has the form
\begin{equation}
\Sigma_{FL}''(\omega) = - \alpha \max\left(\omega^2,T^2\right)
\label{eq:SigmappFL}
\end{equation}
which would be expected to be valid at large doping and low temperatures.  
In the doping region beyond but near the QCP, the single particle properties 
of the model are observed to cross over from FL to MFL as the temperature 
crosses $T_X$ and the frequency $\omega_x$, we find that this may be fit 
with the form
\begin{equation}
\Sigma_{X}''(\omega) = \left\{ \begin{array}{c} 
- \alpha \omega_x \max\left( \left|\omega\right|,T\right) \mbox{ for } |\omega|>\omega_x \mbox{ or } T>T_X \\
- \alpha \max\left(\omega^2,T^2\right) \mbox{ for } |\omega|<\omega_x \mbox{ and } T<T_X 
\end{array} \right.
\label{eq:SigmappX}
\end{equation}
Causality requires that $\alpha>0$, and the integrals over these forms for 
the self energy are cutoff at $\omega_c$ which is of the order of the 
bandwidth $W(=8t)$.

To compare our Matsubara frequency results to these forms of the real-frequency
self energy\cite{HessSerene}, we transform each of these forms to Matsubara 
frequency using the transform of the non-Hartree part of the self energy
\begin{equation}
\Sigma(i\omega_n) = -\int \frac{d\omega}{\pi} \frac{\Sigma''(\omega)}{i\omega_n -\omega}
\end{equation}
and then evaluate 
\begin{equation}
Z_0(\k) = \left(1- {\rm{Im}}\Sigma(\k,i\omega_0)/\omega_0\right)^{-1}
\label{eq:Zmats}
\end{equation}
where $\omega_0=\pi T$ is the lowest Fermion Matsubara frequency.  For a
well behaved self energy, $\lim_{T\to 0} Z_0(\k) = Z(\k)$ is the quasiparticle 
renormalization factor.  For example, for the MFL, we find\cite{HessSerene}
\begin{equation}
\frac{\Sigma_{MFL}(\k,i\omega_0)}{\omega_0} = \frac{\alpha}{\pi} 
\left[ \ln\left(\frac{(\pi^2+1)T^2}{\pi^2T^2+\omega_c^2} \right)
-\frac{2}{\pi}\tan^{-1}\frac1\pi \right]\,.
\label{eq:Z0MFL}
\end{equation}
While for the FL, we find
\begin{equation}
\frac{\Sigma_{FL}(\k,i\omega_0)}{\omega_0} = 
\frac{-2\alpha T}{\pi}\left(
\frac{\omega_c}{T}+0.066235- \pi \tan^{-1}\frac{\omega_c}{\pi T} 
\right)
\label{eq:Z0FL}
\end{equation}
when $T<\omega_c$.  The crossover form is more complicated, but can be 
constructed from the same integrals used to derive Eqs.~\ref{eq:Z0MFL} 
and \ref{eq:Z0FL}  
\begin{flalign}
&\left(-\frac{\pi}{2\alpha}\right)\frac{{\rm Im}\Sigma(i\omega_0)}{\omega_0}
= T\Theta(T_X - T)\Big[\frac{\omega_x}{T} + 0.06623 \notag \\
-& \left(0.308\frac{\omega_x}{\pi T} + \pi\tan^{-1}\frac{\omega_x}{\pi T}\right)
-\frac{\omega_x}{2T}\ln\left(\frac{\omega_x^2+\pi^2T^2}{(1+\pi^2)T^2}\right)\Big]\notag \\
+ \omega_x&\left[0.0981 + \frac{1}{2}\ln\left(\frac{\omega_c^2+\pi^2T^2}
{(1+\pi^2)T^2}\right)\right]\,.
\label{eq:Z0X}
\end{flalign}
The parameters, $\alpha$, $\omega_c$, $T_X$, and $\omega_x$ are determined 
from a fit to the quantum Monte Carlo data.

\paragraph*{Results-} We will present results for the model with $U=1.5$ 
with bare bandwidth $W=2$ setting the energy unit, and a $4\times 4$ 
cluster.  With this choice of $U/t$ and cluster we are able to access low 
temperatures $T \agt 0.01$ before the average sign of the sampling weight 
falls below $0.05$. The low energy scales in the problem are the 
antiferromagnetic exchange energy $J$ near half filling, the PG 
temperature $T^*$ in the PG region, and the effective Fermi energy $T_X$ 
at higher doping.  As described previously\cite{macridin_HEkink}, we can 
extract the effective near neighbor spin exchange energy $J_{eff}$ from the 
spin excitation spectrum at $\K=(0,\pi)$.  At this wavevector the magnon 
peak is expected at frequency $\omega\approx 2J_{eff}$. From this analysis,
we find that $J_{eff}\approx 0.11$ for $N=0.95$ and $N=1$. The energy 
scales $T^*$ and $T_X$ are extracted from fits to the data presented below.  
It is important to note that in each of these fits, we include data for 
$T\ll J_{eff}$.

\begin{figure}[t]
\begin{center}
\includegraphics*[width=3.3in]{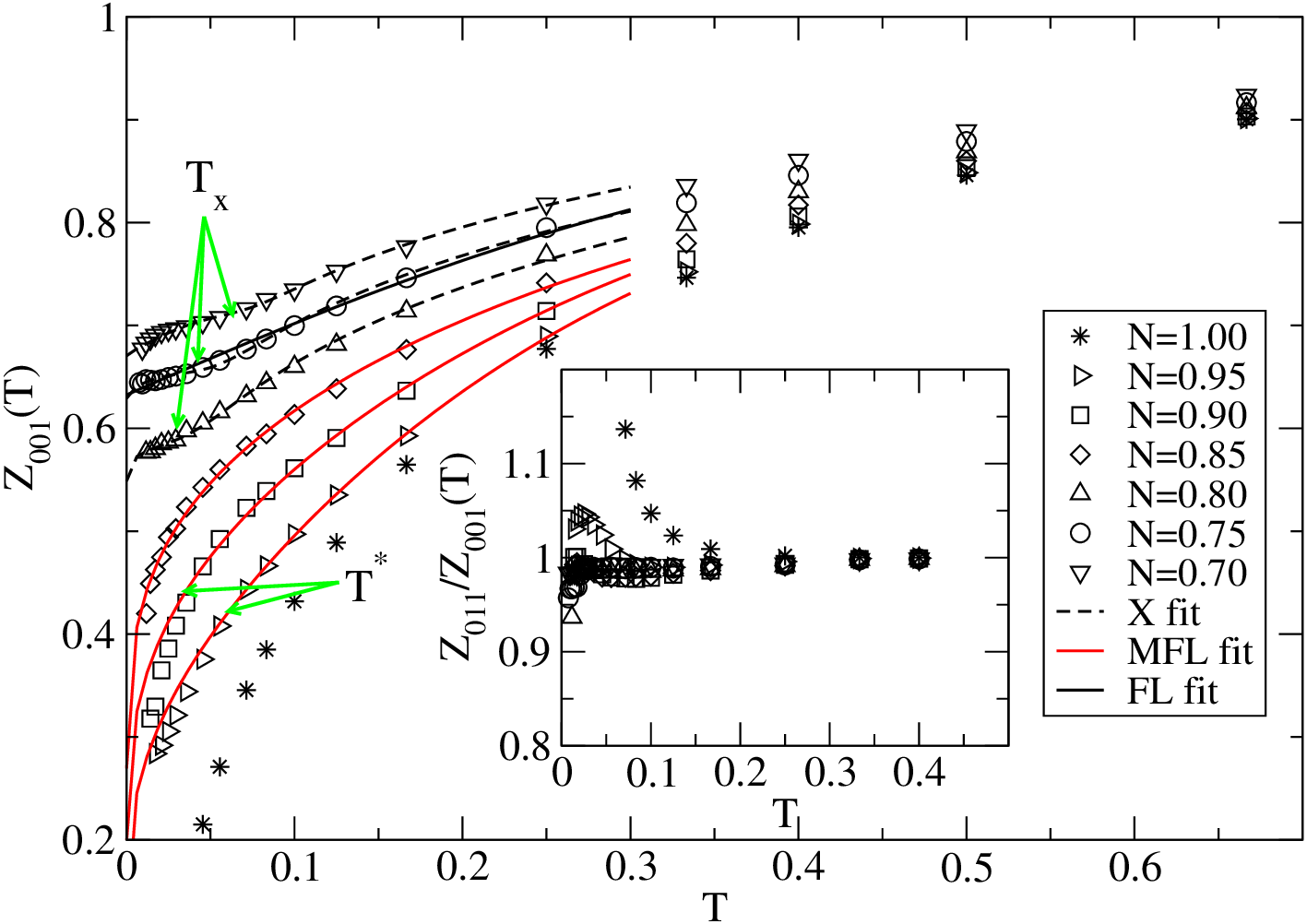}
\caption{
Matsubara quasiparticle fraction $Z_0(\k)$ versus temperature $T$ evaluated
with $\k$ on the Fermi surface along the $(0,1)$ direction for different 
fillings $N$ when $U=1.5$ and the bandwidth  $W=2$.  The lines represent 
fits in the region $T<0.3$ to either the MFL form, Eq.~\ref{eq:Z0MFL}, 
for $N\ge 0.85$ or the crossover form (X), Eq.~\ref{eq:Z0X}, for $N<0.85$.  
The arrows indicate the values $T_X$ extracted from the crossover fits
or $T^*$ (c.f.\ Figs.~\ref{fig:DOS_Nc16B_U1.5tp0_N0.95_vT} and 
\ref{fig:temperatures}).  
Note that the data for $N=0.85$ fits the MFL nearly perfectly, while the 
data for $N > 0.85$ is poorly fit by the MFL for $T<T^*$   
which is too slow in temperature to provide a good fit due to the formation 
of the pseudogap.  The data for $N=0.75$ was also fit by the FL form, 
Eq.~\ref{eq:Z0FL}; however, the fit is clearly worse than that obtained by 
the crossover form. Inset: The ratio, $Z_{011}/Z_{001}$,
is plotted as a function of temperature for different fillings. The ratio
is essentially the same for all fillings at the QCP, indicating that
Z is essentially isotropic, and becomes progressively more
anisotropic as we dope into the PG region.}
\label{fig:Z01_16B_U1.5tp0}
\end{center}
\end{figure}
The Matsubara quasiparticle fraction is calculated with $\k$ on
the Fermi surface defined by the maximum $\left|\nabla n(\k)\right|$ along 
the $(1,1)$ and $(0,1)$ directions.  
We will present detailed results and analysis for the latter
only as we are interested in the crossover from PG to FL behavior, and the 
PG is stronger along the $(0,1)$ direction. $Z_{001}$ is shown in the main panel in
Fig.~\ref{fig:Z01_16B_U1.5tp0} for $U=1.5$ in units where $W=2$ for different 
fillings.  

The low temperature Matsubara quasiparticle data changes character as the 
filling $N$ increases through $N=0.85$.  The data for $N>0.85$ has negative 
curvature at all $T$.  Whereas the data for $N<0.85$ has negative curvature 
at high $T$, a region of weak positive curvature is found at lower $T$.   
The change in curvature of the low temperature data for $N<0.85$ is easily 
understood as a crossover to a FL region.  In a FL at zero 
temperature $Z_{0FL}(0)= 1/(1+2\alpha\omega_c/\pi)$ while for low $T$, 
$Z_{0FL}(T) \approx Z_{0FL}(0)\left( 1+3.099 \alpha Z_{0FL}(0) T\right)$.
Since $\alpha Z_{0FL}(0)>0$, $Z_{0FL}(T)$ at low $T$ has a finite intercept 
and a linear region with positive initial slope indicative of FL formation. 
The next correction, of order $T^2$, is small for $Z_{0FL}(0)\approx 0.6$
so that $Z_{0FL}(T)$ is a nearly linear function when fit to the low $T$ data.
On the other hand, the MFL always has negative curvature, as can be seen 
from an expansion of Eqs.~\ref{eq:Z0MFL} and \ref{eq:Zmats} to second 
order in $T$.  So at the transition between FL and MFL, a region of positive 
curvature is found at $T\approx T_X$.  The data for $N<0.85$ is well fit by 
the crossover form posed above, but is poorly fit by a FL form over the 
fitting region (see e.g., the solid line fit to the $N=0.75$ data).  When 
the data at $N=0.85$ is fit with the crossover form for the $Z_{0X}(T)$, 
the fitting routine returns $\omega_x=T_X=0$ (within the precision of the 
fit and data), consistent with the formation of a MFL.  So the solid line 
shown in the plot is a MFL fit.  The MFL fit to the $N=0.85$ data is very good. 
In fact the quality of this fit was better than that obtained for any of 
the fitting forms to any of the other data sets, despite the fact that the 
MFL form only has two adjustable parameters.
In order to show that the conclusions from the above analysis are not specific
to the direction $(0,1)$, we plot the ratio, $Z_{011}/Z_{001}$, in the inset of
Fig.~\ref{fig:Z01_16B_U1.5tp0} as a function of temperature for different fillings.
The ratio
is seen to be essentially the same for all fillings at the QCP, indicating that
Z is essentially isotropic at the QCP, and becomes progressively more
anisotropic as we dope into the PG region.

\begin{figure}[t]
\begin{center}
\includegraphics*[width=3.3in]{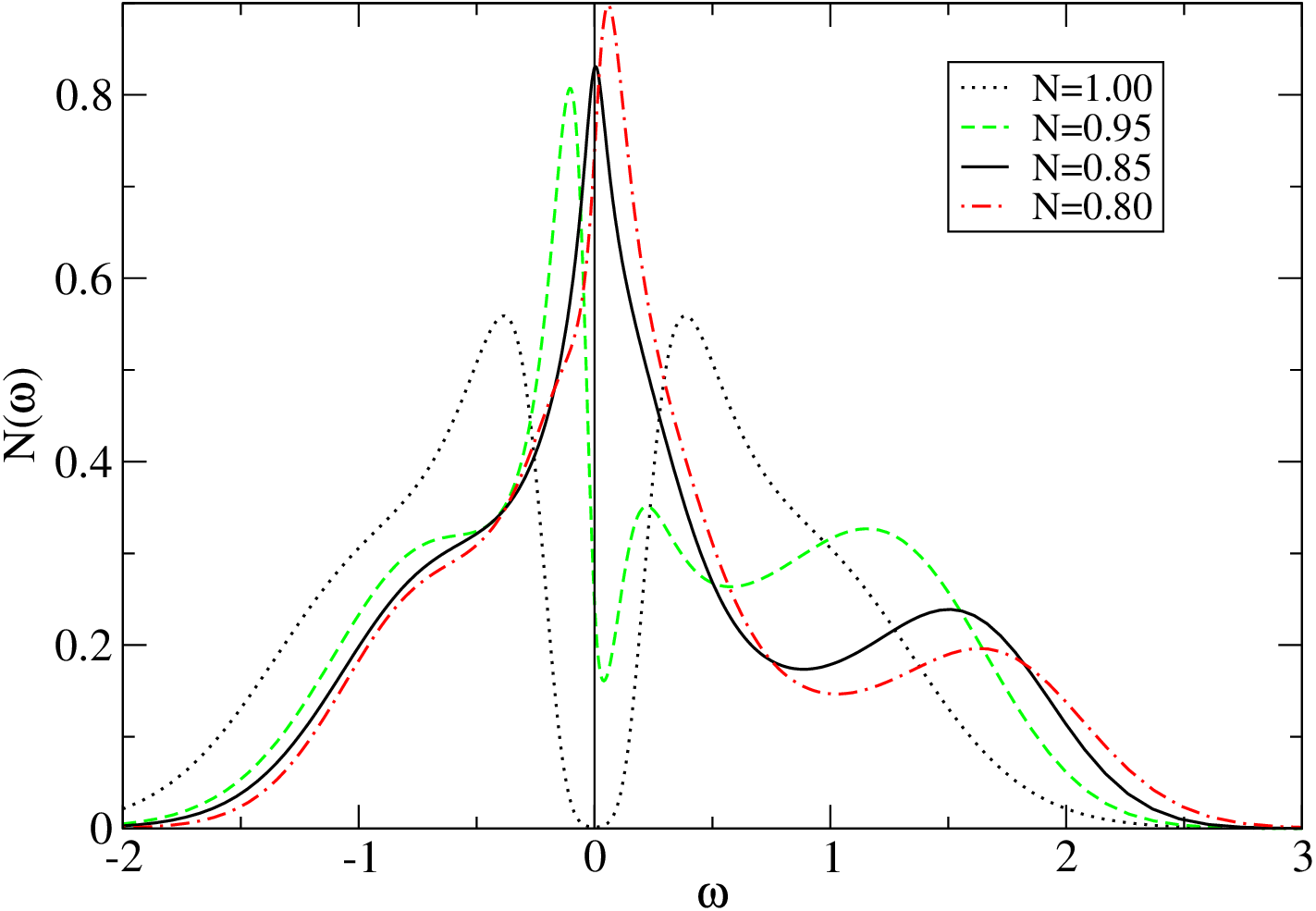}
\caption{
The single-particle density of states (DOS) when $U=1.5$, $W=2$, and 
$T=0.01429$.  Note that the low-energy DOS is roughly symmetric around 
$\omega=0$ for the critical filling $N=0.85$ where $Z_0(\k)$ fits a MFL 
form.  This is consistent with the observation of rough particle-hole  
symmetry in the cuprates in the proximity of optimal doping\cite{s_chakraborty_08}.
 }
\label{fig:DOS_16B_tp0U1.5}
\end{center}
\end{figure}

The data with $N=0.85$ in some other ways is special.  For example, at this
filling the low temperature single-particle density of states (DOS), which
is plotted in Fig.~\ref{fig:DOS_16B_tp0U1.5} for several fillings, is peaked 
at zero frequency. At low energies $|\omega|\alt J_{eff}$ the $N=0.85$
DOS is nearly symmetric around this point. 
This is consistent with the observation of particle-hole (p-h) symmetry in 
the transport of the cuprates at optimal doping\cite{s_chakraborty_08}.

\begin{figure}[t]
\begin{center}
\includegraphics*[width=3.3in]{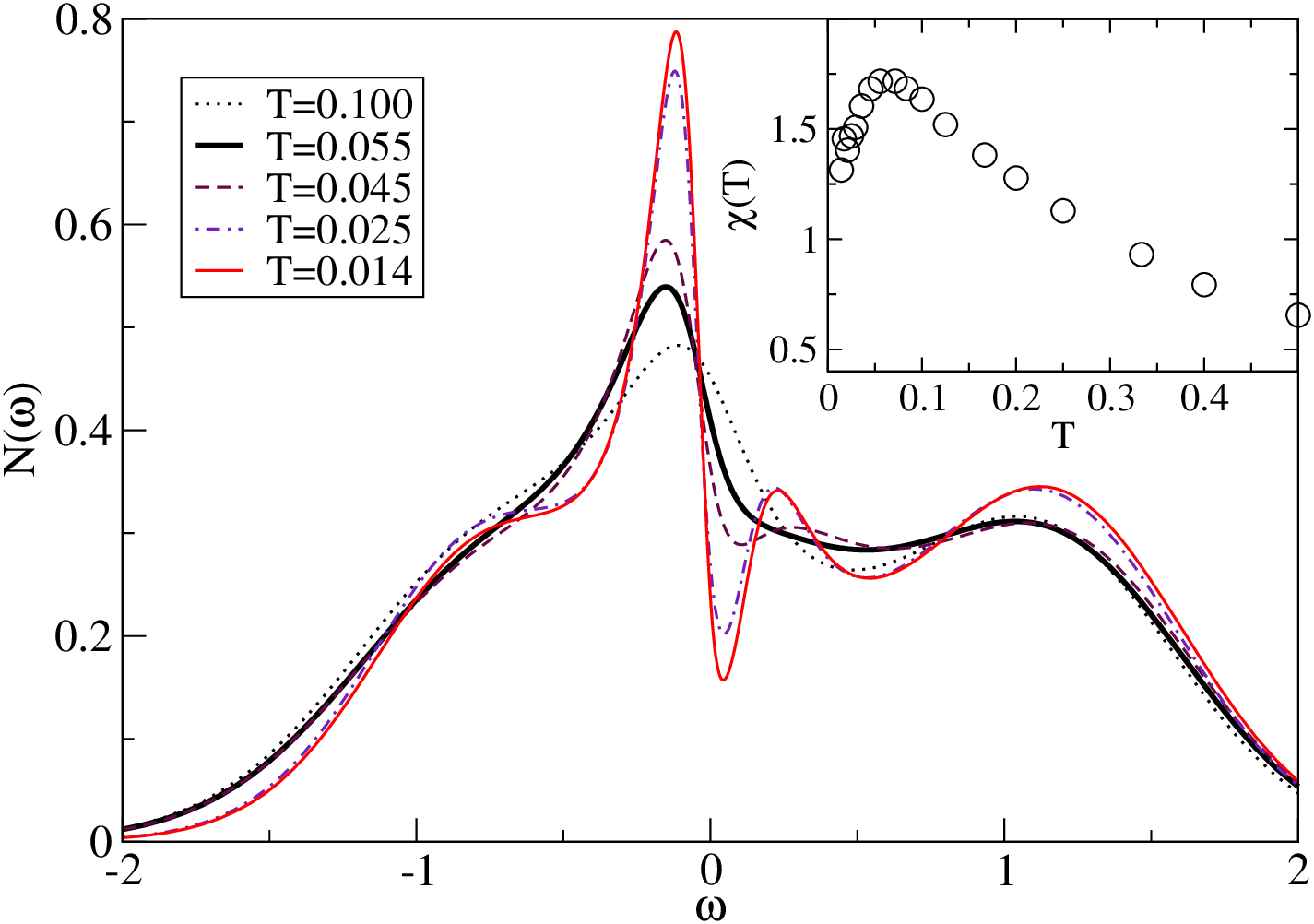}
\caption{
The single-particle density of states in the pseudogap region
for various temperatures with $N=0.95$, $U=1.5$, $W=2$. 
Inset: The bulk, $\Q=0$, cluster susceptibility for the
same parameters.  The PG in the DOS begins to develop at roughly the same 
temperature $T^*$ which identifies the peak susceptibility.
}
\label{fig:DOS_Nc16B_U1.5tp0_N0.95_vT}
\end{center}
\end{figure}

\begin{figure}[t]
\begin{center}
\includegraphics*[width=3.3in]{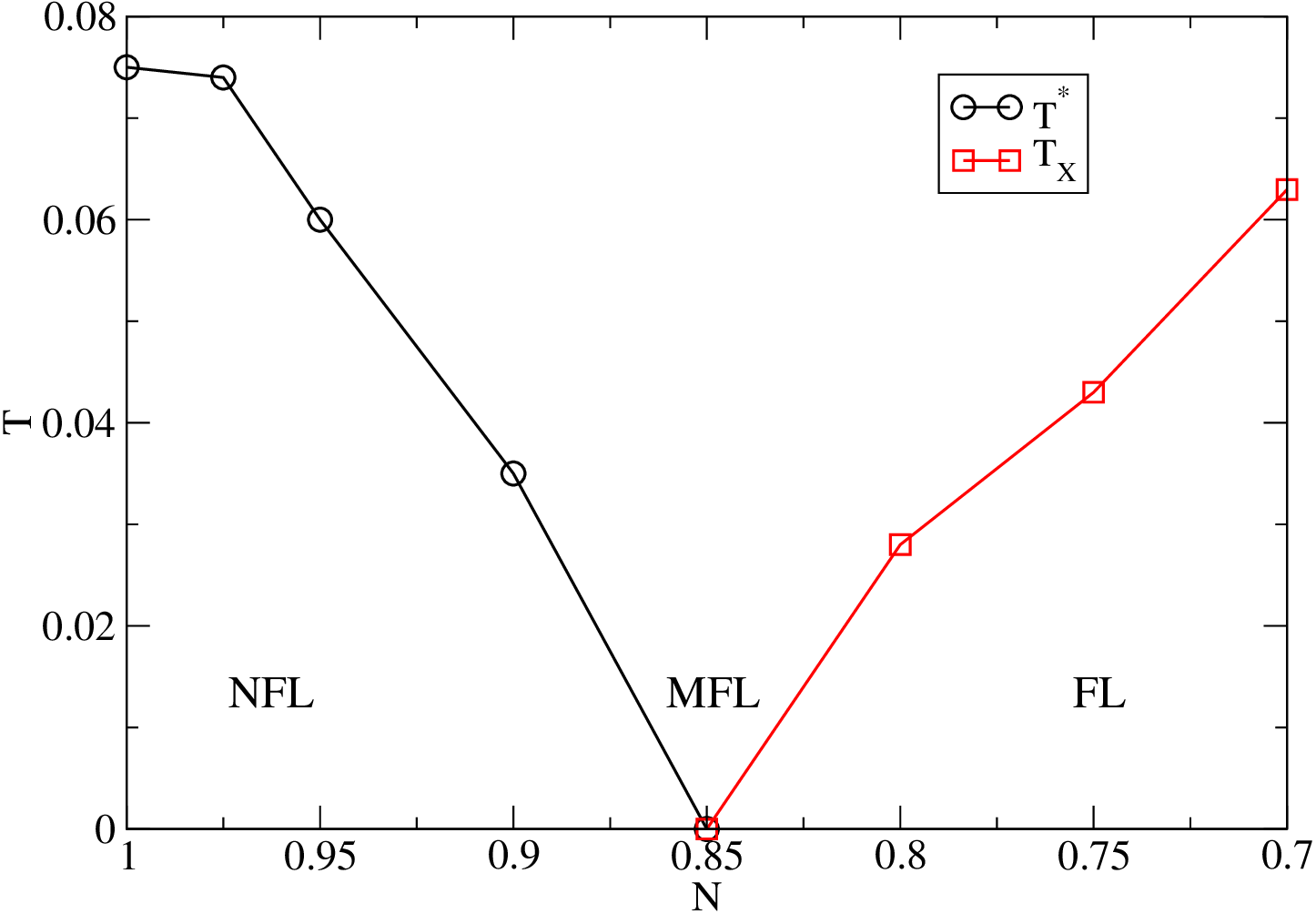}
\caption{
The pseudogap temperature $T^*$, identified from the peak in the 
susceptibility and the emergence of the PG in the DOS shown in 
Fig.~\ref{fig:DOS_Nc16B_U1.5tp0_N0.95_vT} and the FL to MFL 
crossover temperature identified by fitting Eq.~\ref{eq:Z0X} to 
the Matsubara quasiparticle data shown in Fig.~\ref{fig:Z01_16B_U1.5tp0}.
}
\label{fig:temperatures}
\end{center}
\end{figure}

In order to characterize the region $N>0.85$, the PG region, we
also explored the temperature dependence of the DOS and the bulk 
($\Q=0$) spin susceptibility of the cluster, as shown in 
Fig.~\ref{fig:DOS_Nc16B_U1.5tp0_N0.95_vT} and its inset, respectively.
We find a concomitant depression of the low energy DOS at temperatures below 
the peak in the susceptibility.  The suppression of the susceptibility 
indicates the suppression of low energy spin excitations.  $Z_{001}(T)$ 
in this region is well fit for $T>T^*$ (see Fig.~\ref{fig:temperatures}) by 
the MFL form, but is poorly fit by the MFL form for $T<T^*$, as shown in 
Fig.~\ref{fig:Z01_16B_U1.5tp0}.  The MFL form changes too slowly with 
decreasing $T$, due to the formation of the PG for $T<T^*$.

The relevant temperatures near the QCP, $T_X$ and $T^*$, are shown in 
Fig.~\ref{fig:temperatures}.  The PG temperature was determined from
the peak in the susceptibility and the initial appearance of the PG
in the DOS as shown in Fig.~\ref{fig:DOS_Nc16B_U1.5tp0_N0.95_vT},
and $T_X$ from the fit to Eq.~\ref{eq:Z0X}.

We also explored the effect of larger clusters and $U/t$; however, these 
results are restricted by computational limitations including, especially, 
the minus sign problem. Despite this, some calculations were possible for 
$N_c=24$ site clusters.  For the same parameters used above, we find that 
the PG temperature $T^*$ increases slightly near half filling.  The 
critical filling where the $Z_{001}(T)$ is best fit by a MFL, the DOS is 
p-h symmetric at low frequencies remains between $N\approx 0.85$ 
to $N\approx 0.86$.  
Due to the minus sign problem, we did not have enough low
temperature data to determine the crossover temperature precisely for
clusters with $N_c=24$. We also explored larger values of $U=W=2$ for
the $N_c=16$ cluster (this data was produced in an earlier study).  Again 
conclusions are limited by the minus sign problem; however, we find
that the PG vanishes at roughly $N=0.78$, consistent with optimal
doping in the cuprates\cite{s_chakraborty_08}, and at this filling 
the DOS is again p-h symmetric at low energies.

\paragraph*{Discussion-}  Several different scenarios have been proposed to 
explain the transition from PG to FL behavior and the strange behavior seen 
near optimal doping in the cuprates such as the MFL phenomena.   

An extension of the  bond-solid scenario~\cite{Voijta,Sachdev,vojta:cmat08} 
is consistent with our results.  Here, a bond-solid is conjectured to coexist 
with the pseudogap in the underdoped region,
and the QCP marks the doping where the transition 
temperature to bond order vanishes.  We suggest that the NFL behavior could 
be due to the scattering of quasiparticles from bond excitations.  For 
dopings greater than, but near, the PG region, there would be remnant bond 
excitations.  The low energy scale of these excitations 
will be cut off by the finite correlation length of the bond order, yielding 
a gap to bond excitations.  So low energy quasiparticles may form a FL, while 
higher energy ones do not due to scattering from these bond excitations.  In 
this scenario, the gap to bond excitations is proportional to $T_X$, which 
presumably will grow as the bond correlation length falls as the system is 
doped away from the QCP.  One problem with this scenario is that it requires 
gapless bond excitations in the PG region, but due to the finite size of the 
cluster there should be a gap which scales with the cluster size.  However, 
this gap may be small so its effects might not be seen at the temperatures 
we can access.

The spectral-weight transfer scenario\cite{m_meinders_93,s_chakraborty_08} 
also provides a consistent interpretation.  In a hole-doped Mott insulator, 
with doping $x=1-N$ and large $U/t$, each doped hole yields two states 
immediately above the chemical potential.  One comes from the lower and the 
other from the upper Hubbard band.  When the system is doped so that the 
number of low-energy states (i.e., not including those in the upper band) 
above and below the chemical potential are equal, so that $2x=1-x$ or 
$x=1/3$, then p-h symmetry is obtained.  For finite $U/t$, the critical 
doping for p-h symmetry is smaller\cite{s_chakraborty_08}.  This scenario 
also provides a general mechanism for the breakdown of the FL in the PG 
region\cite{p_phillips_08}. In a FL there is a one-to-one correspondence 
between the particles and quasiparticles which means that the number of
ways of adding a particle at low energies equals the number of ways that 
one can add an electron to the unoccupied states. This correspondence breaks 
down in the PG region since the number of ways to add an electron is $2x$, 
while the number of ways one can add a particle is larger\cite{s_chakraborty_08}. 
Since such an argument relies on the asymmetry of adding a hole or an electron, 
it is clearly invalid when p-h symmetry is achieved for the low energy states.
Thus this scenario explains the p-h symmetry of the DOS at the QCP, as 
found in Fig.~\ref{fig:DOS_16B_tp0U1.5}.

Our results differ from previous extended DCA results for the t-J 
model on a $N_c=4$ cluster where a FL-FL crossover, with maximum scattering 
is found at a critical point not associated with MFL behavior\cite{haule}.  
It is not clear whether the model or the method is responsible for these 
differences, although note that the spectral-weight transfer  
arguments discussed above suggest FL physics for the  
t-J model even at small doping\cite{p_phillips_08}.

\paragraph*{Conclusion-}
We investigate the Matsubara quasiparticle fraction 
on the Fermi surface and the PG of the two-dimensional Hubbard model.  
As a function of doping, $Z_{001}(T)$ changes character.  For doping beyond 
the critical point, as the temperature is lowered the curvature of $Z_{001}(T)$ 
changes from negative to positive. This can be understood as a change from 
a MFL to a FL as $T$ falls.  At lower doping, the curvature is negative for 
all $T$, including $T\ll J_{eff}$, consistent with a NFL state.  A PG is also 
found in the DOS and the bulk spin susceptibility at lower doping.  At the 
QCP which separates these two regions, we find a MFL which is also found for 
$T>T^*$ and $T>T_X$.

\acknowledgments  This research was supported by NSF DMR-0312680, DMR-0706379 
(MJ and NSV),  IUSSTF (NSV), DOE CMSN DE-FG02-04ER46129 (AM), and by the 
DOE SciDAC grant DE-FC02-06ER25792 (CS and MJ).  Supercomputer support was
provided by the Ohio Supercomputer Center.  We would like to thank 
R.\ Gass, P.\ Phillips and M.\ Vojta for useful conversations.

\end{document}